\documentstyle[aps]{revtex}
\begin{document}
\draft


\def\bb{\begin{equation}}
\def\ee{\end{equation}}
\title{Polaron transport and lattice dynamics in colossal magnetoresistance manganites}

\author{J. D. Lee and B. I. Min}

\address{Department of Physics, Pohang University of Science and
         Technology,\\ Pohang 790-784, Korea}

\maketitle

\begin{abstract}

Based on the model combining the spin double exchange and the lattice
polaron, we have studied the colossal magnetoresistance  phenomena
observed in perovskite manganites R$_{1-x}$A$_x$MnO$_3$.
First, effects of both the double exchange and the 
electron-phonon interaction on the transport property are investigated.
We have evaluated the temperature dependent resistance and 
the magnetoresistance using the Kubo formula, and examined
the crossover from tunneling to hopping regime of small polarons.
Second, effects of the double exchange
interaction on the lattice degree of freedom are explored.
It is found that both the hardening of the phonon frequency 
and the reduction of the phonon damping take place with 
decreasing the temperature.

\end{abstract}


%
%

\section{Introduction}

The "colossal" magnetoresistance (CMR) manganites R$_{1-x}$A$_x$MnO$_3$ 
(R= La, Pr, Nd; A= Ca, Ba, Sr, Pb) have recently attracted considerable
attention due to scientific interest and potential applicability
of their very large magnetoresistance (MR) for 
$0.2\lesssim x\lesssim 0.5$\cite{Jin,Chahara,Helmolt}. 
The most essential feature of their magnetic and transport behaviors
is the existence of metallic conductivity and ferromagnetism.
The magnetic transition at $T_c$ is closely connected with the resistivity 
peak at $T_P$ corresponding to an insulator-metal transition ($T_c\sim T_P$).
The correlation between ferromagnetism and metallic
conductivity in R$_{1-x}$A$_x$MnO$_3$ was explained 
by Zener\cite{Zener} in terms of the
double exchange mechanism. 
There are mixed valent Mn ions (Mn$^{3+}$ and Mn$^{4+}$)
as a consequence of hole doping by substituting R$^{3+}$ with A$^{2+}$. 
In the double exchange model,
conduction electrons in the partially filled $e_g$ levels of the $d$ band are
strongly coupled with the tightly bound $d$ electrons in the $t_{2g}$ levels
by the on-site Hund's coupling, and mediate the ferromagnetic 
exchange interaction between the nearest neighbor $S=\frac{3}{2}$ local spins 
formed from three $d$ electrons in the core-like
$t_{2g}$ levels\cite{Zener,Anderson}.

Transport properties have been studied 
within the double exchange mechanism in favor of a magnetic 
polaron\cite{Kubo,Kusters,Furukawa}. Recently, 
Millis {\it et al.}\cite{Millis} reported that the
effective carrier-spin interaction involved in the
ordinary double exchange Hamiltonian is too weak to
produce the magnetic polaron effects. Instead,
they suggested lattice polaron effects 
due to a strong electron-phonon interaction 
as a necessary additional extension \cite{Millis2}. 
They investigated a model of electrons, Jahn-Teller coupled to localized
classical oscillators, within the dynamical mean field theory.
R\"{o}der {\it et al.}\cite{Roder} also examined the combined
influence of the electron-phonon interaction and the double exchange
on $T_c$ using the variational wave function techniques.
But they could not treat the polaron transport and 
the lattice dynamics on an equal footing.
In fact, the contribution of the lattice polaron to carrier mobility 
was pointed out earlier by Goodenough \cite{Goodenough}.

There are many experimental evidences
suggesting importance of the electron-lattice coupling
in manganese oxides\cite{Hwang,Ibarra,Khkim,Jeong,Ram}.
Near $T_c$, dramatic changes are observed in the
lattice degree of freedom  - the anomalous lattice expansion
beyond Gr\"{u}neisen law\cite{Ibarra}, and the shift of phonon frequency 
\cite{Khkim,Jeong,Ram}, which all reflect that the lattice 
is closely related to the electronic and magnetic properties.
However, detailed understanding of the interplay between 
the lattice dynamics and the electronic and magnetic properties 
remains to be resolved. 

In this paper, we have addressed two questions;
i) what is the role of the electron-phonon interaction in
CMR systems which are known to have the double exchange interaction, and reversely,
ii) how the double exchange interaction affects the lattice dynamics
through the electron-phonon interaction. 
For these purposes, we first investigate effects of
both the double exchange and the electron-phonon interaction on  
transport and magnetic properties.
Employing the Kubo formula, we have determined 
the temperature dependent resistance and the magnetoresistance,
and examined the crossover from a metallic tunneling state
to an insulating hopping state in the small polaron transport.
Second, to characterize the lattice dynamics in CMR compounds,
we have considered the phonon degree of freedom in the presence of
the double exchange interaction. We have studied the
renormalization of the phonon frequency and the phonon damping constant.

This paper is organized as follows.
In section II, we present the model of conduction electrons
coupled to phonons as well as the localized ionic spins in terms of
the double exchange, whereby
the temperature dependent resistance and the magnetoresistance are
evaluated from the Kubo formula. In section III, we examine the
double exchange effects on the lattice degree of freedom.
Finally, conclusions follow in section IV.
Detailed calculational steps are given in Appendix.

%
%

\section{Polaron transport}

Since Zener\cite{Zener} has proposed an interaction between
spins of magnetic ions named "double exchange",
Anderson and Hasegawa\cite{Anderson} studied this mechanism
in a system of Mn ions and a mobile electron with the transfer
$t$ between two Mn ions and the strong intra-atomic exchange
integral $J$. When $J$ is much larger than $t$,
motion of the mobile electrons in R$_{1-x}$A$_x$MnO$_3$ is described 
by the following double exchange Hamiltonian,
\bb
{\cal H}_{\rm DE}=\sum_{ij}t_{ij}cos\frac{\theta_{ij}}{2}
                  c_i^{\dagger}c_j,
\ee
where the hopping $t_{ij}$ connects neighboring sites, and
$\theta_{ij}$ is the angle between the directions of ionic spins
at sites $i$ and $j$. An exact quantum mechanical
calculation gives
$
cos\frac{\theta_{ij}}{2}=\frac{S_0+1/2}{2S+1},
$
where $S$ is the spin of a Mn ion and $S_0$ is the total spin of
$S_i$, $S_j$ and the conduction electron spin.
In this study, we treat the double exchange part within the
mean field theory following Kubo and Ohata\cite{Kubo},
in which $cos\frac{\theta_{ij}}{2}$ is replaced by its thermodynamic
average $\langle cos\frac{\theta_{ij}}{2}\rangle$
determined by minimizing the free energy of the spin system.
Then the propagation of an electron can be described
as if it were moving in a mean field of highly disordered
configurations of ionic spins.
This approximation is known to work well at finite temperature,
except for the very low temperature region $(T\sim 0$K) where the
spin dynamics becomes important.
Within the present mean field theory,
the double exchange plays a role, through 
$\langle\frac{S_0+1/2}{2S+1}\rangle\equiv\gamma(T)$, of increasing the
bandwidth as $T$ decreases below $T_c$, accompanied by the ferromagnetic
ordering (see Fig.~\ref{bandw}(a)).

In addition to the double exchange, the conduction electrons are 
scattered by the Mn-O ionic motions in the MnO$_6$ octahedra, which
gives rise to a very strong electron-phonon interaction. 
The effective Hamiltonian incorporating
the electron-phonon interaction is written as,
\bb
{\cal H}=t\langle cos\frac{\theta}{2}\rangle\sum_{i\delta}
         c_{i+\delta}^{\dagger}c_i
        +\sum_{\vec{q}}\omega_q a_{\vec{q}}^{\dagger}a_{\vec{q}}
        +\sum_{i\vec{q}}c_i^{\dagger}c_i e^{i\vec{q}\cdot\vec{R}_i}
         M_q(a_{\vec{q}}+a_{-\vec{q}}^{\dagger}).
\ee 
Here we adopt a model in which the single $e_g$ orbital is coupled
to phonons assuming the electronically active $e_g$ band 
to be split, as in R\"{o}der {\it et al.}'s \cite{Roder}.
The present assumption is expected to be more effective
if the model were generalized to include another physics such as the
on-site Coulomb interaction, which might remove possible mid-gap states
away from the Fermi level\cite{Millis2}.

The dc conductivity $\sigma$ can be obtained from the optical 
conductivity $\sigma(\omega)$ by taking the $\omega\rightarrow 0$ limit,
and $\sigma(\omega)$ can be determined by using the 
Kubo formula of the current-current correlation function, 
\bb
\sigma(\omega)=\frac{1-e^{-\beta\omega}}{2\omega}
               \int_{-\infty}^{\infty}d\tau e^{i\omega\tau}
               \langle J_{\alpha}^{\dagger}(\tau)J_{\alpha}(0)\rangle.
\ee
Since the current operator $\vec{J}$ in narrow band systems is given by
\bb
\vec{J}=it\langle cos\frac{\theta}{2}\rangle{\rm e}\sum_{j\delta}
        \hat{\delta}c_{j+\delta}^{\dagger}c_j,
\ee
$\sigma$ explicitly involves the four-site correlation function,
\bb
\sigma=\frac{\beta}{2}t^2\langle cos\frac{\theta}{2}\rangle^2{\rm e}^2
       \sum_{\delta\delta^{\prime}}\sum_{jj^{\prime}}
       (\hat{\delta}\cdot\hat{\delta^{\prime}})
       \int_{-\infty}^{\infty}d\tau
       \langle c_j^{\dagger}(\tau)c_{j+\delta}(\tau)
       c_{j^{\prime}+\delta^{\prime}}^{\dagger}c_{j^{\prime}}\rangle.
\ee
In the isotropic case, the resistivity $\rho$ corresponds to 
the inverse of $\sigma$, $\rho=1/\sigma$.

To evaluate $\sigma$, let's consider the well-known 
polaron canonical transformation\cite{Mahan};
$
\bar{{\cal H}}=e^S{\cal H}e^{-S}
$
with
$
S=-\sum_{j\vec{q}}c_j^{\dagger}c_j e^{i\vec{q}\cdot\vec{R}_j}
   \frac{M_q}{\omega_q}(a_{\vec{q}}-a_{-\vec{q}}^{\dagger}).
$
The transformed Hamiltonian  $\bar{{\cal H}}$ is given by
\bb
\bar{{\cal H}}=t\langle cos\frac{\theta}{2}\rangle\sum_{j\delta}
               c_{j+\delta}^{\dagger}c_j X_{j+\delta}^{\dagger}X_j
              +\sum_{\vec{q}}\omega_q a_{\vec{q}}^{\dagger}a_{\vec{q}}
              -\Delta\sum_j c_j^{\dagger}c_j,
\ee
with 
$
X_j=\exp\left[\sum_{\vec{q}}e^{i\vec{q}\cdot\vec{R}_j}
    \frac{M_q}{\omega_q}(a_{\vec{q}}-a_{-\vec{q}}^{\dagger})\right]
$
and
$\Delta=\sum_{\vec{q}}\frac{M_q^2}{\omega_q}$.
Inserting $e^S e^{-S}=1$ between each electron operator in Eq.(5), 
and using $e^S c_j e^{-S}=c_j X_j$ and
$e^S c_j^{\dagger}e^{-S}=c_j^{\dagger}X_j^{\dagger}$,
one gets
\bb
\sigma=\frac{\beta}{2}t^2\langle cos\frac{\theta}{2}\rangle^2{\rm e}^2
       \sum_{\delta\delta^{\prime}}\sum_{jj^{\prime}}
       (\hat{\delta}\cdot\hat{\delta^{\prime}})
       \int_{-\infty}^{\infty}d\tau
       \langle c_j^{\dagger}(\tau)c_{j+\delta}(\tau)
       c_{j^{\prime}+\delta^{\prime}}^{\dagger}c_{j^{\prime}}
       X_j^{\dagger}(\tau)X_{j+\delta}(\tau)
       X_{j^{\prime}+\delta^{\prime}}^{\dagger}X_{j^{\prime}}\rangle.
\ee

The intricate correlation function
of Eq.(7) should be evaluated under the transformed Hamiltonian
$\bar{{\cal H}}$.
Calculation can be further simplified with
an approximation replacing the first term of Eq.(6) by
$
t\langle cos\frac{\theta}{2}\rangle\sum_{j\delta}
\langle X_{j+\delta}^{\dagger}X_j\rangle c_{j+\delta}^{\dagger}c_j.
$
This approximation is quite reasonable in view of that 
the coherent band-like motion of the polarons arises from
the quantum mechanical tunneling
between sites without changing the phonon numbers, which
is governed by the matrix elements of 
$\langle n_{\bf q}|X_{j+\delta}^{\dagger}X_j|n_{\bf q}\rangle$\cite{Alex}.
With increasing $T$, the polaron bandwidth decreases exponentially
due to the term 
$\langle X_{j+\delta}^{\dagger}X_j\rangle(\equiv \Gamma(T))$,
\bb
\Gamma(T)=\exp[-\sum_{\vec{q}}|u_{\vec{q}}|^2(N_q+1/2)],
\ee
where $u_{\vec{q}}\equiv(M_q/\omega_q)(e^{i\vec{q}\cdot\vec{\delta}}-1)$
(see Fig.~\ref{bandw}(b)).
Under the above approximation, the mean field scheme of 
Kubo and Ohata\cite{Kubo} can be generalized to include 
the phonon contributions which drastically reduce the magnetic
transition temperature $T_c$\cite{Roder}, whereas the temperature
dependent behavior of 
$\langle cos\frac{\theta}{2}\rangle(\equiv\gamma(T))$
does not appreciably change. 

With the approximate Hamiltonian $\bar{{\cal H}}$
incorporating both $\gamma(T)$ and $\Gamma(T)$, 
the complicated four-site correlation function in Eq.(7) 
can be disentangled into 
$\langle c_j^{\dagger}(\tau)c_{j+\delta}(\tau)
 c_{j^{\prime}+\delta^{\prime}}^{\dagger}c_{j^{\prime}}\rangle
\langle X_j^{\dagger}(\tau)X_{j+\delta}(\tau)
 X_{j^{\prime}+\delta^{\prime}}^{\dagger}X_{j^{\prime}}\rangle$.
These correlation functions 
and $\sigma$ can be evaluated in the straightforward fashion.
Detailed calculational procedures are provided in Appendix.
From Eq.(A12), the dc conductivity $\sigma$ is given as follows,
\begin{eqnarray}
\sigma&=&\frac{\beta}{2}t^2\gamma(T)^2{\rm e}^2
       \sum_{\delta\delta^{\prime}}\sum_{jj^{\prime}}
       (\hat{\delta}\cdot\hat{\delta}^{\prime})
       \sum_{\vec{k}_1\vec{k}_2}
       n_{\vec{k}_1}(1-n_{\vec{k}_2})
       e^{-i\vec{k}_1\cdot(\vec{R}_j-\vec{R}_{j^{\prime}})}
       e^{i\vec{k}_2\cdot(\vec{R}_j-\vec{R}_{j^{\prime}}
                    +\vec{\delta}-\vec{\delta}^{\prime})} \nonumber \\
      &\times&
       e^{(\tilde{t}_{\vec{k}_1}-\tilde{t}_{\vec{k}_2})/2T}
       e^{-(\tilde{t}_{\vec{k}_1}-\tilde{t}_{\vec{k}_2})^2
           /4\zeta(j,j^{\prime},\vec{\delta},\vec{\delta}^{\prime};T)}
       e^{-\xi(T)}
       e^{\eta(j,j^{\prime},\vec{\delta},\vec{\delta}^{\prime};T)}
       [\pi/\zeta(j,j^{\prime},\vec{\delta},\vec{\delta}^{\prime};T)]^{1/2},
\end{eqnarray}
with a renormalized polaron band,
$\tilde{t}_{\vec{k}}=t\gamma(T)
\Gamma(T)\sum_{\delta}e^{-i\vec{k}\cdot\vec{\delta}}-\Delta$.
Explicit expressions of 
$\zeta(j,j^{\prime},\vec{\delta},\vec{\delta}^{\prime};T)$,
$\xi(T)$, and $\eta(j,j^{\prime},\vec{\delta},\vec{\delta}^{\prime};T)$
are given in Appendix.
Keeping in mind that the auto-correlation function is most dominant
for $j=j^{\prime}$ and $\vec{\delta}=\vec{\delta}^{\prime}$,
$\sigma$ is more simply obtained in the following form
\bb
\sigma=\frac{\beta}{2}t^2\gamma(T)^2{\rm e}^2 Nz
       \sum_{\vec{k}_1\vec{k}_2}
       n_{\vec{k}_1}(1-n_{\vec{k}_2})
       e^{(\tilde{t}_{\vec{k}_1}-\tilde{t}_{\vec{k}_2})/2T}
       e^{-(\tilde{t}_{\vec{k}_1}-\tilde{t}_{\vec{k}_2})^2/4\zeta(T)}
       e^{-\xi(T)}e^{\eta(T)}(\pi/\zeta(T))^{1/2}.
\ee
Here $z$ is the number of the nearest neighbors, 
$n_{\vec{k}}$ and $N_q$ are the fermion and boson distribution function,
respectively, and $\xi(T)$, $\zeta(T)$, and $\eta(T)$ are also 
given by
\bb
\xi(T)=\sum_{\vec{q}}|u_{\vec{q}}|^2(1+2N_q),
\ee
\bb
\zeta(T)=\sum_{\vec{q}}\omega_q^2|u_{\vec{q}}|^2[N_q(N_q+1)]^{1/2},
\ee
\bb
\eta(T)=2\sum_{\vec{q}}|u_{\vec{q}}|^2[N_q(N_q+1)]^{1/2}.
\ee

Now let's investigate the qualitative behavior of $\rho$.
One can carry out the numerical calculation of Eq.(10),
assuming the simple square density of states (DOS) ${\cal D}(\epsilon)$, 
\bb
{\cal D}(\epsilon)=\frac{N(1-x)}{\epsilon_F},\mbox{ }\mbox{ }
-\frac{W}{2}\leq\epsilon\leq \frac{W}{2}.
\ee
Here the bandwidth $W$ is given by $w\gamma(T)\Gamma(T)$
($w$ is the bare electron bandwidth and 
estimated to be $w=12|t|\sim 2{\rm eV}$ from the band structure calculation),
and the Fermi energy $\epsilon_F$ is $w\gamma(T)\Gamma(T)(1-x)$
with $x$ being the doping concentration. 
These treatments of DOS are based on the assumption that
only the lower one of the split $e_g$ bands is active.
We also assume that the most relevant phonon mode is 
the optical mode ($\sim\omega_0$) which might be involved in
the Jahn-Teller coupling.

Numerical results for the resistivity are provided in Fig.~\ref{res}.
Two dashed lines represent resistivities of the polaron only model
with given bandwidths of $\tilde{t}_{\rm up} (=w\Gamma(T))$ and 
$\tilde{t}_{\rm low} (=0.75w\Gamma(T))$.
Here $\tilde{t}_{\rm up}$ and $\tilde{t}_{\rm low}$
correspond, respectively, to upper and lower limits of
the double exchange factor $\gamma(T)$. 
In both cases, $\rho$'s exhibit peaks as a function of $T$,
and the larger band width $\tilde{t}_{\rm up}$ yields a smaller $\rho$ 
with a peak at higher temperature.
The resistivity peak as a function of $T$ corresponds to the crossover from 
a quantum tunneling of the metallic phase to
a self-trapped small polaron hopping of the insulating phase.
Such features are characteristics of polaron systems, which are
indeed observed in many oxide systems.
In the high $T$ limit, the resistivity has 
a thermal activation form of $\exp(\Delta_g/T)$, characteristic of
the semiconducting phase\cite{Mahan}.  
Now taking the double exchange into account, the polaron bandwidth
increases with decreasing $T$ due to $\gamma(T)$, and accordingly the
resistivity is given by the solid line (in Fig.~\ref{res}) 
with a peak at $T_c$ connecting
the two polaron resistivity curves.
This figure clearly demonstrates that the semiconducting behavior 
above $T_c$ is attributed to self-trapped lattice small polarons, 
and that the rapid fall-off in the resistivity 
below $T_c$ is attributed to the double exchange mechanism in addition to
the lattice polaron effect.
Thus, the combined model of the double exchange and the polaron provides
a good description of the resistance anomaly observed in the experiment.
The coincidence of the resistivity peak position $T_P$ with
$T_c$ originates from the mean field treatment of $\gamma(T)$
which neglects the fluctuation in the hopping of conduction electrons.

Effects of the external magnetic field can also be taken into account in
$\gamma(T)$ through the modified free energy due to the magnetic field.
The behaviors of the MR's are shown in the inset of Fig.~\ref{res}. 
With increasing the field intensity, the resistivity decreases and
the peak position shifts to a higher $T$, and so the negative MR results.
These results are quite consistent with the experimental observations. 
The MR peak $T_{MR}$ is located near the resistivity peak $T_P$.
In fact, $T_{MR}$, $T_P$, and $T_c$ are the same in the present
mean field treatment.
It should be noticed that the magnitudes of the MR's in the figure
are not large enough to explain the experimental CMR data 
quantitatively, suggesting that additional treatments might
be required. 
One possibility is to incorporate the half-metallic 
nature of the ferromagnetic manganites\cite{Pickett,Ysj}, which
is expected to suppress largely the spin-disorder scattering 
under the external magnetic field.

%
%

\section{Lattice dynamics}

As mentioned before, the phonon frequency 
becomes  hardened as $T$ decreases below $T_c$\cite{Khkim,Jeong,Ram}. 
Interestingly, the phonon hardenings are observed for 
both optical and acoustic phonons in these systems.
These frequency shifts are considered to be due to the change in 
the electron screening as $T$ is lowered below $T_c$. 
The hardening occurs in the metallic region ($T\lesssim T_c$), {\it i.e.},
in the band-type tunneling regime where Rayleigh-Schr\"{o}dinger 
perturbation theory is valid\cite{Mahan}. 
Therefore it is expected that the change of the
bandwidth due to the double exchange factor $\gamma(T)$
modifies the electron screening below $T_c$.
Note that the previous approximation for the canonically transformed 
Hamiltonian $\bar{{\cal H}}$ corresponds
to neglecting the phonon frequency renormalization,
which seems to be too small to cause any appreciable change in the
transport properties.  In the metallic regime, 
the screening of the conduction electrons will be more easily 
described by the original Hamiltonian ${\cal H}$ of Eq.(2) rather than
the polaron Hamiltonian $\bar{{\cal H}}$ of Eq.(6). 

The renormalized phonon frequency 
$\tilde{\omega}_q$ and the damping constants $\alpha_q$ 
can be obtained from the following equation\cite{Djkim}
\bb
(\tilde{\omega}_q-i\alpha_q)^2=\omega_q^2-2\omega_q|M_q|^2
                             {\cal F}(\vec{q}, \tilde{\omega}_q+i0^+),
\mbox{ }\mbox{ }
\tilde{\omega}_q \gg \alpha_q,
\ee
where the electron screening function 
${\cal F}(\vec{q}, \tilde{\omega}_q)$ is given by
\bb
{\cal F}(\vec{q}, \tilde{\omega}_q)
=\sum_{\vec{q}}\frac{n_{\vec{k}}-n_{\vec{k}+\vec{q}}}
               {\gamma(T)(t_{\vec{k}+\vec{q}}-t_{\vec{k}})-
                \tilde{\omega}_q},
\mbox{ }\mbox{ }
t_{\vec{k}}=t\sum_{\delta}e^{i\vec{k}\cdot\vec{\delta}}.
\ee
The real part of both sides of Eq.(15) yields
\bb
\tilde{\omega}_q^2\simeq\omega_q^2-2\omega_q|M_q|^2\frac{1}{\gamma(T)}
         \sum_{\vec{k}}\frac{n_{\vec{k}}-n_{\vec{k}+\vec{q}}}
                       {t_{\vec{k}+\vec{q}}-t_{\vec{k}}}.
\ee 
It is important to note that the term
$\sum_{\vec{q}}(n_{\vec{k}}-n_{\vec{k}+\vec{q}})
/(t_{\vec{k}+\vec{q}}-t_{\vec{k}})$
has very weak temperature dependences ($\sim C + {\cal O}(T/E_F)^2$, 
$E_F$ being the Fermi level determined from $t_{\vec{k}}$).
Hence the $T$-dependence of $\tilde{\omega}_q$
comes dominantly from $\gamma(T)$, and Eq.(17) can be written
as $\tilde{\omega}_q=\omega_q(1-\bar{\beta}/\gamma(T))^{1/2}$, where 
$\bar{\beta}$ contains all the $T$-independent terms.
An explicit calculation of $\bar{\beta}$ is not available, 
but the order of its magnitude should be very small, 
${\cal O}(|M_q|^2/(\omega_q E_F)) \sim{\cal O}(10^{-2})$.
In Fig.~\ref{phn}(a), $T$-dependent behaviors of $\tilde{\omega}_q$ are plotted
with respect to the external magnetic field strength $H$,
and compared with the available experiment\cite{Jeong} in the inset. 
It is seen that the frequency hardenings with decreasing $T$ and with
increasing $H$ are qualitatively well explained.
However, it is also apparent that some deviations exist between calculational
and experimental results, particularly, near $T_c$. 
These discrepancies might be ascribed to the mean field 
treatment of $\gamma(T)$.
Including spin correlation effects in calculating $\gamma(T)$
is expected to improve the agreement. 
It should also be noted that the phonon frequency hardening 
of Eq.(17) would be valid for both acoustic and optical phonon modes
consistently with the experiments,
because we have assumed the general form of the electron-phonon
interaction in Eq.(2).

Taking the imaginary part for   
both sides of Eq.(15), one gets the phonon damping parameter $\alpha_q$,
\bb
2\alpha_q\tilde{\omega}_q
      =2\omega_q|M_q|^2\frac{1}{\gamma(T)}{\rm Im}
       \sum_{\vec{k}}\frac{n_{\vec{k}}-n_{\vec{k}+\vec{q}}}
       {t_{\vec{k}+\vec{q}}-t_{\vec{k}}-\tilde{\omega}_q/\gamma(T)-i0^+}.
\ee 
The imaginary part of the screening function is easily
calculated by considering a parabolic electron 
band $t_{\vec{k}}=t\sum_{\delta}e^{i\vec{k}\cdot\vec{\delta}}
\simeq |t|\delta^2k^2$, 
\bb
2\alpha_q=2\omega_q|M_q|^2\frac{\pi}{2}
        \frac{{\cal D}(E_F)}{v_F q}
        \frac{1}{\gamma(T)^2},
\ee
where ${\cal D}(E_F)$ and $v_F$ are determined from the parabolic band, 
$|t|\delta^2k^2$.
In Fig.~\ref{phn}(b), $T$-dependent behaviors of $\alpha_q$ are presented.
Our results predict that the phonon damping parameter decreases with
decreasing $T$, implying that the phonon is more sharply defined
below $T_c$. This feature in CMR systems is quite different from 
conventional observations of increased phonon damping parameter below $T_c$
for magnetic or strongly correlated systems.
To our knowledge, no experimental reports are 
available yet on the phonon damping parameter. 
We think that the sound attenuation experiment will provide a better
understanding of the nature of the electron-phonon interaction in CMR systems.

%
%

\section{Conclusions}

We have extended the double exchange model 
to incorporate the strong electron-phonon interaction, 
and investigated transport and
magnetic properties of CMR manganese oxides R$_{1-x}$A$_x$MnO$_3$.
We have found that the semiconducting behavior
in manganites above $T_c$ is attributed to the
effect of self-trapped lattice small polarons, 
and that the rapid fall-off in the resistivity
below $T_c$ is attributed to the combined effect of 
coherent lattice polarons and the increased bandwidth
via the double exchange mechanism accompanied by
the ferromagnetic ordering of magnetic ions.
Further, we have explored effects of the double
exchange on the phonon degrees of freedom. The temperature 
dependent hardening of the phonon mode frequency observed in experiments
is well described, and the reduction of the phonon damping constant
is predicted below $T_c$.

%
%

\acknowledgements

This work was supported by the Korea Research Foundation, 
and in part by the BSRI program of the
Korean Ministry of Education and the POSTECH special fund.
Helpful discussions with Y.H. Jeong and T.W. Noh are greatly appreciated.

%
%

\appendix
\section*{Calculation of \lowercase{dc} conductivity $\sigma$}

To evaluate the dc conductivity $\sigma$, one should evaluate
correlation functions of electrons and phonons under the Hamiltonian $\bar{{\cal H}}$.
Defining simply $\Lambda$ and $\Lambda^{\prime}$ as
\bb
\Lambda=e^{i\vec{q}\cdot\vec{R}_j}(e^{i\vec{q}\cdot\vec{\delta}}-1)
        \frac{M_q}{\omega_q},
\mbox{ }\mbox{ }
\Lambda^{\prime}=e^{i\vec{q}\cdot\vec{R}_j^{\prime}}
        (e^{i\vec{q}\cdot\vec{\delta}^{\prime}}-1)\frac{M_q}{\omega_q},
\ee
$X_j^{\dagger}(\tau)X_{j+\delta}(\tau)
X_{j^{\prime}+\delta^{\prime}}^{\dagger}X_{j^{\prime}}$ is given by
\bb
X_j^{\dagger}(\tau)X_{j+\delta}(\tau)
X_{j^{\prime}+\delta^{\prime}}^{\dagger}X_{j^{\prime}}=
\prod_{\vec{q}}e^{-\frac{1}{2}(|\Lambda|^2+|\Lambda^{\prime}|^2)}
e^{\Lambda^{\ast} a_{\vec{q}}^{\dagger}e^{i\omega_q\tau}}
e^{-\Lambda a_{\vec{q}}e^{-i\omega_q\tau}}
e^{-{\Lambda^{\prime}}^{\ast}a_{\vec{q}}^{\dagger}}
e^{\Lambda^{\prime}a_{\vec{q}}}.
\ee
Using $e^{-\Lambda a_{\vec{q}}e^{-i\omega_q\tau}}
e^{-{\Lambda^{\prime}}^{\ast}a_{\vec{q}}^{\dagger}}
=e^{-{\Lambda^{\prime}}^{\ast}a_{\vec{q}}^{\dagger}}
 e^{-\Lambda a_{\vec{q}}e^{-i\omega_q\tau}}
 e^{\Lambda{\Lambda^{\prime}}^{\ast}e^{-i\omega_q\tau}}$,
we see
\bb
X_j^{\dagger}(\tau)X_{j+\delta}(\tau)
X_{j^{\prime}+\delta^{\prime}}^{\dagger}X_{j^{\prime}}=
\prod_{\vec{q}}e^{-\frac{1}{2}(|\Lambda|^2+|\Lambda^{\prime}|^2)}
e^{\Lambda{\Lambda^{\prime}}^{\ast}e^{-i\omega_q\tau}}
e^{\lambda^{\ast}a_{\vec{q}}^{\dagger}}
e^{-\lambda a_{\vec{q}}},
\ee
where $\lambda\equiv\Lambda e^{-i\omega_q\tau}-\Lambda^{\prime}$.
Under the noninteracting phonon Hamiltonian,
the thermodynamic average of
$\langle e^{\lambda^{\ast}a_{\vec{q}}^{\dagger}}e^{-\lambda a_{\vec{q}}}
\rangle$ is given by\cite{Mahan}
\bb
\langle e^{\lambda^{\ast}a_{\vec{q}}^{\dagger}}e^{-\lambda a_{\vec{q}}}\rangle
=e^{-|\lambda|^2 N_q},
\mbox{ }\mbox{ }
N_q=\frac{1}{e^{\beta\omega_q}-1}.
\ee
Then $\langle X_j^{\dagger}(\tau)X_{j+\delta}(\tau)
X_{j^{\prime}+\delta^{\prime}}^{\dagger}X_{j^{\prime}}\rangle$ is obtained
as follows
\bb
\langle X_j^{\dagger}(\tau)X_{j+\delta}(\tau)
X_{j^{\prime}+\delta^{\prime}}^{\dagger}X_{j^{\prime}}\rangle
=\exp[-\Phi(\vec{R}_j-\vec{R}_{j^{\prime}},\vec{\delta},
                   \vec{\delta}^{\prime},\tau)],
\ee
\begin{eqnarray}
\Phi(\vec{R}_j-\vec{R}_{j^{\prime}},\vec{\delta},\vec{\delta}^{\prime},\tau)
&=& \sum_{\vec{q}}|u_{\vec{q}}|^2(1+2N_q) \nonumber \\
&-& 2\sum_{\vec{q}}v_{\vec{q}}(j,\vec{\delta})
          v_{\vec{q}}^{\ast}(j^{\prime},\vec{\delta}^{\prime})
         [N_q(N_q+1)]^{1/2}cos[\omega_q(\tau+i\frac{\beta}{2})],
\end{eqnarray}
where $u_{\vec{q}}\equiv(M_q/\omega_q)(e^{i\vec{q}\cdot\vec{\delta}}-1)$ and
$v_{\vec{q}}(j,\vec{\delta})\equiv(M_q/\omega_q)
 e^{i\vec{q}\cdot\vec{R}_j}(e^{i\vec{q}\cdot\vec{\delta}}-1)$.

The electron four-site correlation function
$\langle c_j^{\dagger}(\tau)c_{j+\delta}(\tau)
c_{j^{\prime}+\delta^{\prime}}^{\dagger}c_{j^{\prime}}\rangle$
is easily evaluated from the Hamiltonian $\bar{{\cal H}}$
which transformed into the $\vec{k}$-space,
\bb
\langle c_j^{\dagger}(\tau)c_{j+\delta}(\tau)
c_{j^{\prime}+\delta^{\prime}}^{\dagger}c_{j^{\prime}}\rangle
        =\sum_{\vec{k}_1,\vec{k}_2}n_{\vec{k}_1}(1-n_{\vec{k}_2})
         e^{i(\tilde{t}_{\vec{k}_1}-\tilde{t}_{\vec{k}_2})\tau}
         e^{-i(\vec{k}_1-\vec{k}_2)\cdot(\vec{R}_j-\vec{R}_{j^{\prime}})}
         e^{i\vec{k}_2\cdot(\vec{\delta}-\vec{\delta}^{\prime})},
\ee
where the renormalized polaron band $\tilde{t}_{\vec{k}}$ is given as 
$t\gamma(T)\langle X_{j+\delta}^{\dagger}
X_j\rangle\sum_{\delta}e^{i\vec{k}\cdot\vec{\delta}}$.

The time integral of the correlation functions, Eq.(7) can be performed
by the saddle point approximation.
In the vicinity of the saddle 
point, the integrand becomes just a Gaussian,
\bb
\Phi(\vec{R}_j-\vec{R}_{j^{\prime}},\vec{\delta},\vec{\delta}^{\prime},\tau)
\simeq\xi(T)-\eta(j,j^{\prime},\vec{\delta},\vec{\delta}^{\prime};T)
+\zeta(j,j^{\prime},\vec{\delta},\vec{\delta}^{\prime};T)z^2,
\mbox{ }\mbox{ }
z=\tau+i\beta/2,
\ee
where $\xi(T)$, $\eta(j,j^{\prime},\vec{\delta},\vec{\delta}^{\prime};T)$,
and $\zeta(j,j^{\prime},\vec{\delta},\vec{\delta}^{\prime};T)$ are,
respectively, given by
\bb
\xi(T)=\sum_{\vec{q}}|u_{\vec{q}}|^2(1+2N_q),
\ee
\bb
\eta(j,j^{\prime},\vec{\delta},\vec{\delta}^{\prime};T)
 =2\sum_{\vec{q}}v_{\vec{q}}(j,\vec{\delta})
  v_{\vec{q}}^{\ast}(j^{\prime},\vec{\delta}^{\prime})
  [N_q(N_q+1)]^{1/2},
\ee
\bb
\zeta(j,j^{\prime},\vec{\delta},\vec{\delta}^{\prime};T)
 =\sum_{\vec{q}}\omega_q^2v_{\vec{q}}(j,\vec{\delta})
  v_{\vec{q}}^{\ast}(j^{\prime},\vec{\delta}^{\prime})
  [N_q(N_q+1)]^{1/2}.
\ee
>From Eqs.(7), (A5), and (A7), the dc conductivity $\sigma$ is
evaluated and obtained as
\begin{eqnarray}
\sigma &=&\frac{\beta}{2}t^2\gamma(T)^2{\rm e}^2
       \sum_{\delta\delta^{\prime}}\sum_{jj^{\prime}}
       (\hat{\delta}\cdot\hat{\delta}^{\prime})
       \sum_{\vec{k}_1\vec{k}_2}
       n_{\vec{k}_1}(1-n_{\vec{k}_2})
       e^{-i\vec{k}_1\cdot(\vec{R}_j-\vec{R}_{j^{\prime}})}
       e^{i\vec{k}_2\cdot(\vec{R}_j-\vec{R}_{j^{\prime}}
                    +\vec{\delta}-\vec{\delta}^{\prime})} \nonumber \\
      &\times&
       e^{(\tilde{t}_{\vec{k}_1}-\tilde{t}_{\vec{k}_2})/2T}
       e^{-(\tilde{t}_{\vec{k}_1}-\tilde{t}_{\vec{k}_2})^2
           /4\zeta(j,j^{\prime},\vec{\delta},\vec{\delta}^{\prime};T)}
       e^{-\xi(T)}
       e^{\eta(j,j^{\prime},\vec{\delta},\vec{\delta}^{\prime};T)}
       [\pi/\zeta(j,j^{\prime},\vec{\delta},\vec{\delta}^{\prime};T)]^{1/2}.
\end{eqnarray}

\begin{figure}
\caption{Bandwidths (hopping parameters in unit of $t$) as 
a function of the temperature.
(a) Effect of the double exchange interaction, 
$t\gamma(T), (\gamma(T)\equiv\langle cos\frac{\theta}{2}\rangle)$.
(b) Combined effect of the double exchange and the electron-phonon
interaction, $t\gamma(T)\Gamma(T)$. $\sum_{\vec{q}}|u_{\vec{q}}|^2=4.5$
is taken.
}
\label{bandw}
\end{figure}

\begin{figure}
\caption{The resistivity of the polaron only model 
(two dashed lines) and the combined
model of the polaron and double exchange (solid line). 
In the polaron only model,
$\tilde{t}_{\rm up}$ and $\tilde{t}_{\rm low}$ are the bandwidths 
corresponding to the the upper and lower limiting value of 
$\langle cos\frac{\theta}{2}\rangle$, respectively.
Calculations are performed for
the parameters, 
$x=0.3$, $w=64T_c\sim 1.1{\rm eV}$, $\omega_0=5T_c\sim 0.08{\rm eV}$
assuming $T_c\sim 200{\rm K}$,  
and $\sum_{\vec{q}}|u_{\vec{q}}|^2=6$.
The inset presents the resistance behaviors with $H=0$T, $4.8$T,
and $9.6$T.
}
\label{res}
\end{figure}

\begin{figure}
\caption{(a) The phonon frequency shifts $(\tilde{\omega}_q(T)-
\tilde{\omega}_q^c)/\tilde{\omega}_q^c$ for various magnetic field strengths, 
where $\tilde{\omega}_q^c\equiv\tilde{\omega}_q(1.1T_c,H=0{\rm T})$.
In the inset, the shifts are  compared with the experiment (under $H=1$T) for 
La$_{0.7}$Ca$_{0.3}$MnO$_3$(Jeong {\it et al.}[16]).
In the fitting, $T_c=238{\rm K}$ 
is taken from experiments. We have taken $\bar{\beta}=0.07$. 
(b) The phonon damping constants $\alpha_q(T)/\alpha_q^c$
with $\alpha_q^c\equiv\alpha_q(1.1T_c,H=0{\rm T})$,
are given. 
}
\label{phn}
\end{figure}

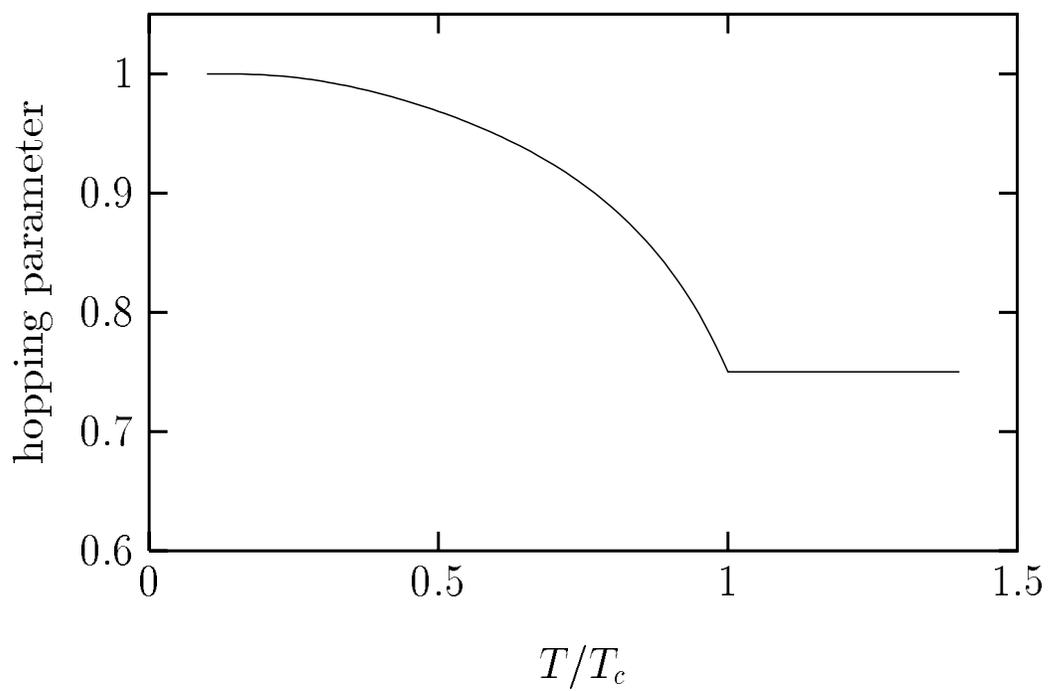
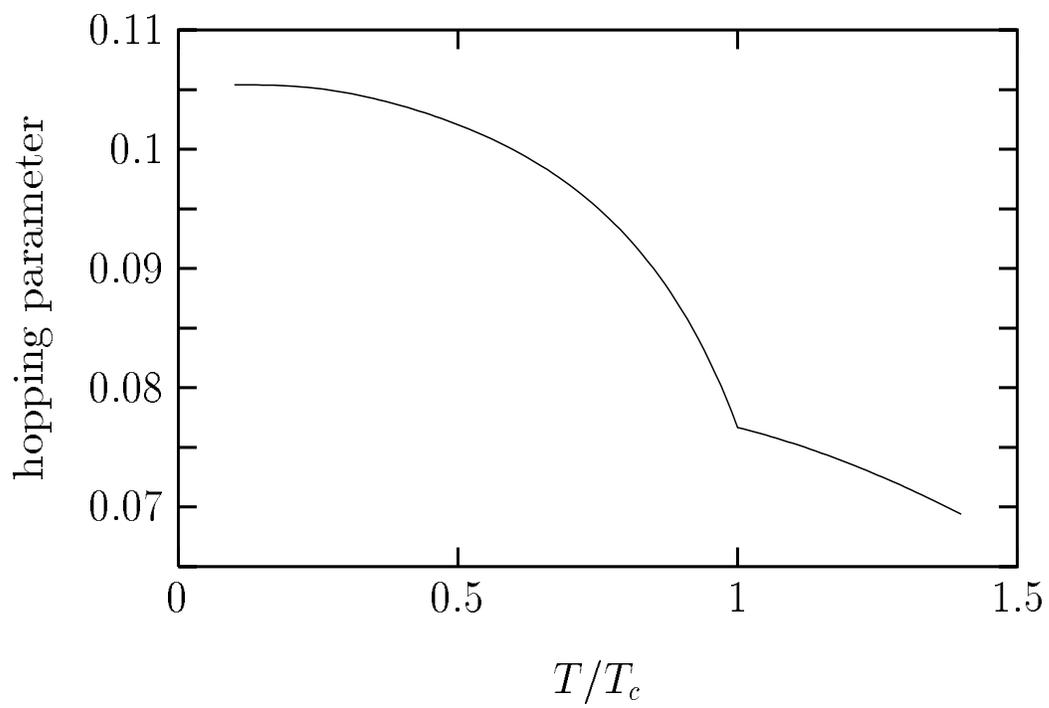
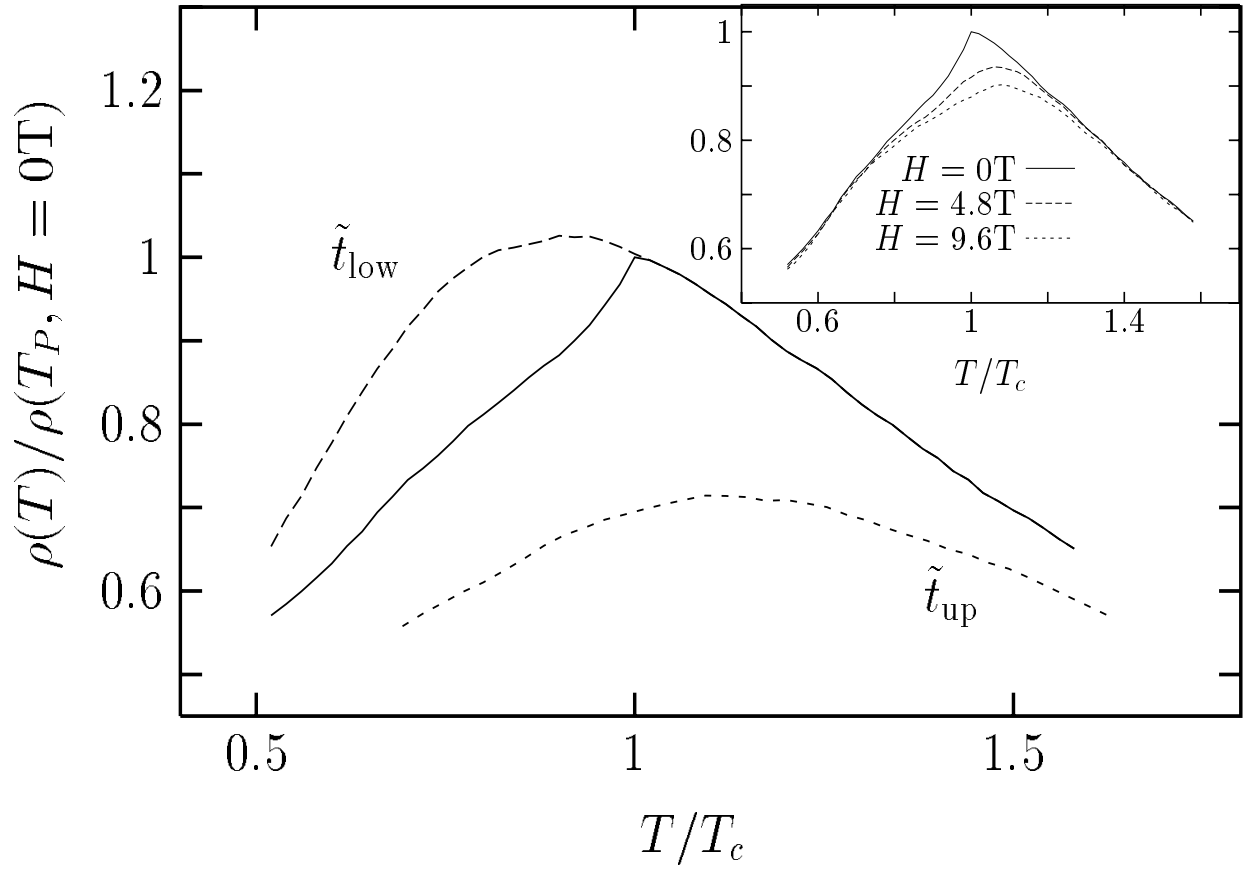
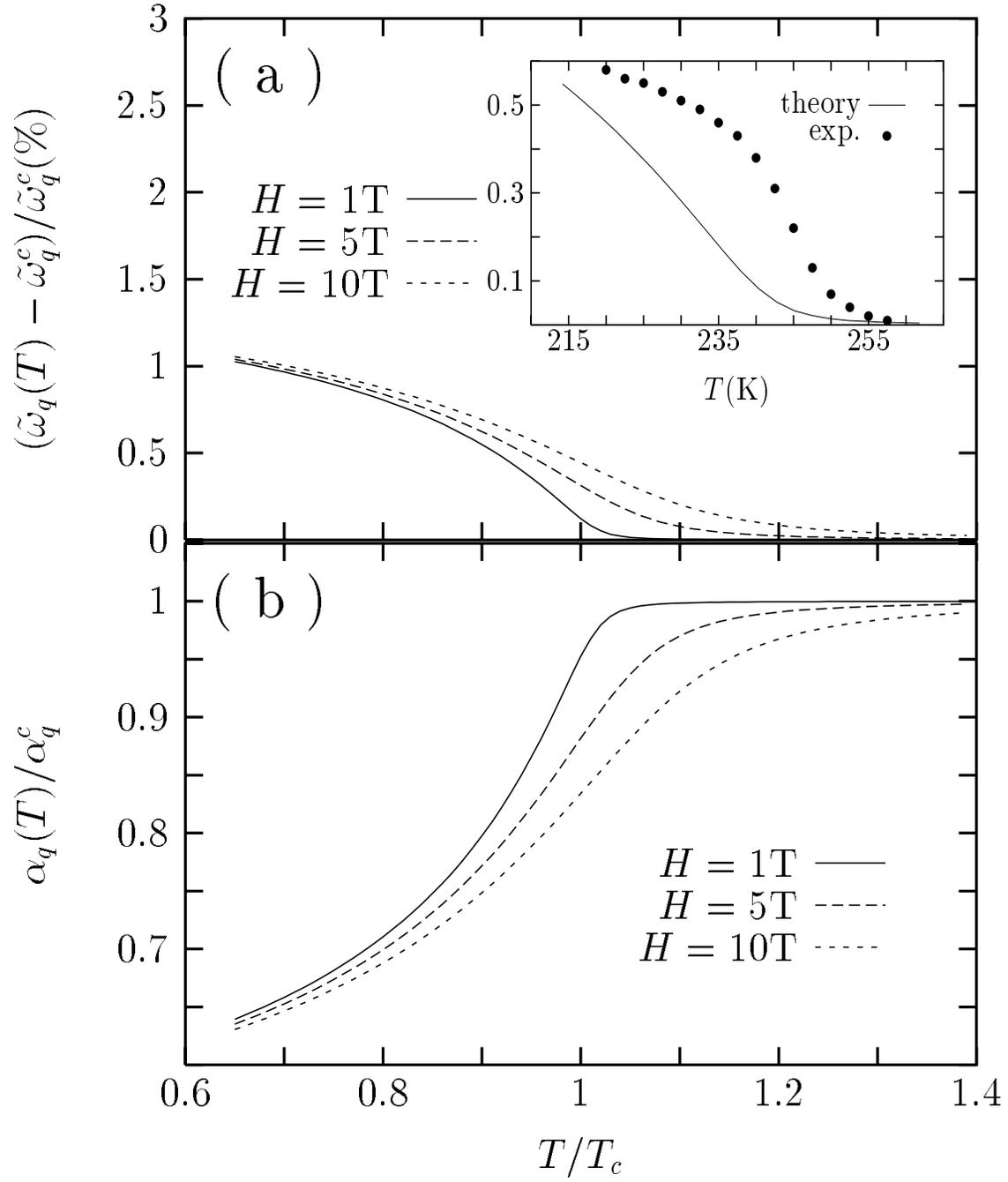
\end{document}